\documentclass[conference]{IEEEtran}
\IEEEoverridecommandlockouts
\usepackage{cite}
\usepackage{amsmath,amssymb,amsfonts}
\usepackage{algorithmic}
\usepackage{graphicx}
\usepackage{textcomp}
\usepackage{xcolor}
\def\BibTeX{{\rm B\kern-.05em{\sc i\kern-.025em b}\kern-.08em
    T\kern-.1667em\lower.7ex\hbox{E}\kern-.125emX}}

\usepackage{url}
\usepackage{siunitx}
\usepackage{booktabs}

\begin{document}

\title{Direction-Preserving MIMO Speech Enhancement Using a Neural Covariance Estimator}

\author{\IEEEauthorblockN{Thomas Deppisch}
\IEEEauthorblockA{\textit{Chalmers University of Technology} \\
Gothenburg, Sweden \\
thomas.deppisch@chalmers.se}
}

\maketitle

\begin{abstract}
Multichannel speech enhancement is widely used as a front-end in microphone array processing systems. While most existing approaches produce a single enhanced signal, direction-preserving multiple-input multiple-output (MIMO) methods instead aim to provide enhanced multichannel signals that retain directional properties, enabling downstream applications such as beamforming, binaural rendering, and direction-of-arrival estimation. In this work, we propose a fully blind, direction-preserving MIMO speech enhancement method based on neural estimation of the spatial noise covariance matrix. A lightweight OnlineSpatialNet estimates a scale-normalized Cholesky factor of the frequency-domain noise covariance, which is combined with a direction-preserving MIMO Wiener filter to enhance speech while preserving the spatial characteristics of both target and residual noise. In contrast to prior approaches relying on oracle information or mask-based covariance estimation for single-output systems, the proposed method directly targets accurate multichannel covariance estimation with low computational complexity. Experimental results show improved speech enhancement, covariance estimation capability, and performance in downstream tasks over a mask-based baseline, approaching oracle performance with significantly fewer parameters and computational cost.
\end{abstract}


\section{Introduction}\label{sec:introduction}
Microphone array–based speech enhancement is a key building block in modern audio systems and hearing aids, enabling robust operation in noisy and reverberant environments. Most common are multiple-input single-output (MISO) approaches that estimate a single enhanced output signal, for example using beamforming or multichannel Wiener filtering~\cite{Souden2010}. While effective for tasks such as speech communication or recognition, this paradigm limits the flexibility of subsequent processing stages that rely on directional characteristics of the sound field. In contrast, direction-preserving MIMO speech enhancement aims to produce enhanced multichannel signals in which both target and residual components retain their spatial structure~\cite{Herzog2019e}. This is particularly beneficial for applications such as binaural rendering and spatial audio, and also supports downstream tasks like adaptive beamforming and direction-of-arrival estimation.

Despite its importance, direction-preserving MIMO speech enhancement has received limited attention. Early formulations~\cite{Herzog2019e} provide a MIMO Wiener filtering framework in the Ambisonics domain, but rely on strong assumptions such as oracle knowledge of the frequency-domain noise covariance. Extensions operate directly on microphone signals, yet still depend on oracle knowledge of spatial noise coherence and reverberation time~\cite{Herzog2021a}. Other works study specific aspects, such as masking-based approaches or covariance structures of a desired source, often under idealized conditions or with oracle information~\cite{Lugasi2020,Lugasi2024a}. A more recent learning-based approach demonstrates spatially consistent multichannel enhancement and source separation, but is restricted to Ambisonics representations and computational costly~\cite{Herzog2023}.

In this work, we instead aim to provide a learning-based direction-preserving MIMO speech enhancement method that directly operates on microphone signals, does not require oracle knowledge, and limits the computational expense of the neural network by reducing the learning task to the estimation of a frequency-domain noise covariance.
Building on~\cite{Herzog2021a}, we employ a direction-preserving MIMO Wiener filter and replace the oracle components with a neural covariance estimator. 

Existing neural covariance estimators for speech enhancement are typically based on single- or multichannel masks. Single-channel mask–based methods estimate a shared mask across microphones to isolate target or noise components and derive covariance matrices for beamforming~\cite{Heymann2016a,Erdogan2016}. While lightweight and robust, they may be limited in spatial and temporal modeling capacity. Extensions incorporate recurrent or attention-based modules to capture temporal dynamics~\cite{Casebeer2021a,Wang2024c}, but remain focused on single-output enhancement, leaving the spatial quality of the estimated covariance for MIMO use unclear. Multichannel mask–based methods estimate channel-dependent masks and obtain covariances via temporal averaging~\cite{Grinstein2025}. Also, alternative non-mask approaches have been proposed~\cite{Pandey2024a}, however, these methods still collapse the spatial dimension and are trained for single-output objectives.

This contribution targets spatial covariance estimation for multichannel output to enable direction-preserving MIMO enhancement. We employ the OnlineSpatialNet architecture~\cite{Quan2024a}, which is specifically designed to model spatial information in online multichannel audio tasks, and train it to estimate a frequency-domain noise covariance. We compare against the mask-based neural integrated covariance estimator (NICE)~\cite{Casebeer2021a} under the same training setup to assess potential advantages of a network architecture that is specifically designed to model spatio-temporal statistics. Exploiting the model-based direction-preserving MIMO Wiener filter for stability, interpretability, robustness, and explicit control of distortion, the proposed approach achieves effective MIMO speech enhancement while preserving directional information with reduced computational complexity.

\section{MIMO Speech Enhancement}
Let $\mathbf{x}(t,f) \in \mathbb{C}^{M}$ denote the multichannel microphone signal in the short-time Fourier transform (STFT) domain with $M$ microphones. The signal is modeled as the sum of a desired speech component $\mathbf{s}(t,f)$ and additive noise $\mathbf{n}(t,f)$,
\begin{equation}
\mathbf{x}(t,f) = \mathbf{s}(t,f) + \mathbf{n}(t,f).
\end{equation}

Assuming the signals are zero-mean and mutually uncorrelated, the spatial covariance matrix of the array observations is~\cite{Benesty2008}
\begin{equation}\label{eq:cov_sum}
\mathbf{R}_{xx}(f)
=
\mathbb{E}[\mathbf{x}\mathbf{x}^\mathrm{H}]
=
\mathbf{R}_{ss}(f) + \mathbf{R}_{nn}(f),
\end{equation}
where $\mathbf{R}_{ss}$ and $\mathbf{R}_{nn}$ are the spatial covariance matrices of speech and noise.
The goal of MIMO speech enhancement is to estimate an enhanced output signal
\begin{equation}
\mathbf{y}(t,f) = \mathbf{W}(t,f)\mathbf{x}(t,f),
\end{equation}
where $\mathbf{W}(t,f) \in \mathbb{C}^{M \times M}$ is a time-varying linear MIMO filter. 
In this work, we consider reverberant speech and noise sources and aim to suppress noise while preserving the spatio-spectral characteristics of the reverberant target to enable natural binaural rendering of the enhanced sound scene.
Throughout the following, the frequency and time indices are omitted for readability.

\section{Direction-Preserving MIMO Wiener Filter}
The MIMO Multichannel Wiener filter (MWF) minimizes the mean squared error~\cite{Doclo2002b}
\begin{equation}
\mathbb{E}\left[\|\mathbf{s}-\mathbf{W}\mathbf{x}\|^2\right]
\end{equation}
and is given by
\begin{equation}
\mathbf{W}_{\mathrm{MWF}}
=
\mathbf{R}_{ss}\mathbf{R}_{xx}^{-1}.
\end{equation}
Using~\eqref{eq:cov_sum}, this can be written as
\begin{equation}\label{eq:W_MWF}
\mathbf{W}_{\mathrm{MWF}}
=
\mathbf I - \mathbf{R}_{nn}\mathbf{R}_{xx}^{-1}.
\end{equation}
While the MIMO MWF provides a solution for multichannel speech enhancement, it may distort the target signal. The parametric MIMO MWF~\cite{Herzog2021a} is a MIMO extension of the parametric MWF~\cite{Spriet2004} that provides a tradeoff between speech distortion and noise reduction. It preserves spatial properties of the target but distorts the spatial characteristics of the residual noise~\cite{Herzog2020}. 

To reduce spatial distortions of the residual noise, the direction-preserving MIMO MWF (DP-MWF) uses a signal-dependent mixture of the Wiener filter and the identity matrix~\cite{Herzog2021a},
\begin{equation}\label{eq:DP_MWF}
\mathbf{W}_\mathrm{DP-MWF} = (1-a')\mathbf{W}_{\mu+\nu} + a' \mathbf{I},
\end{equation}
where $a' \in [0,1]$ is an analytically determined mixing factor,
\begin{equation}
a' = a + (1-a)\frac{\nu \, \mathrm{tr}\!\left(\mathbf{W}_{\mu+\nu}\mathbf{R}_{nn}\right)}
{\mu \, \mathrm{tr}\!\left(\mathbf{R}_{nn}\right) + \nu \, \mathrm{tr}\!\left(\mathbf{W}_{\mu+\nu}\mathbf{R}_{nn}\right)},
\end{equation}
and $a \in [0,1]$ defines a lower bound on the identity mixing.

The Wiener filter component 
\begin{equation}
\mathbf{W}_{\mu+\nu}
=
(\mathbf{R}_{xx}-\mathbf{R}_{nn})
\left(
\mathbf{R}_{xx}+(\mu+\nu-1)\mathbf{R}_{nn}
\right)^{-1}.
\end{equation}
is similar to the MIMO MWF in~\eqref{eq:W_MWF} but additionally controls the trade-off between noise reduction and speech distortion with the parameter $\mu \geq 0$, and the strength of the direction-preserving term with $\nu \geq 0$.

\section{Neural Covariance Estimation}
The DP-MWF has been shown to effectively reduce noise while preserving the spatial properties of both speech and residual noise, but in its original formulation it relies on prior knowledge of the spatial noise coherence matrix and the reverberation time~\cite{Herzog2021a}. In this work, we instead address the fully blind problem by estimating the noise covariance $\mathbf{R}_{nn}$ using a neural network. To limit computational complexity, the mixture covariance $\mathbf{R}_{xx}$ is obtained via causal averaging of the sample covariance over a sliding window.

\subsection{Estimation as Scale-Normalized Cholesky Factor}\label{sec:cholesky-estim}
Before feeding the mixture signal to the neural network, we estimate a frequency-dependent scaling factor based on the average trace of the covariance matrix to reduce sensitivity to the absolute signal level,
\begin{equation}
\gamma(f)
=
\frac{1}{T}
\sum_{t=1}^{T}
\frac{1}{M}\operatorname{tr}\!\left(\mathbf{\hat{R}}_{xx}(t,f)\right).
\end{equation}
Given the normalized mixture STFT,
\begin{equation}
\tilde{\mathbf{x}}(t,f)
=
\frac{\mathbf{x}(t,f)}{\sqrt{\gamma(f)}},
\end{equation}
the network predicts a lower-triangular Cholesky matrix factor
\begin{equation}
\mathbf{L}(t,f)
=
\mathrm{NeuralNetwork}\!\left(\tilde{\mathbf{x}}(t,f)\right).
\end{equation}
The off-diagonal entries are estimated as unconstrained complex values, whereas the diagonal entries are constrained to be real and positive using a softplus nonlinearity with a small floor $\epsilon$,
to ensure that the resulting covariance estimate is Hermitian positive definite.
The scale-normalized noise covariance is then obtained as
\begin{equation}
\tilde{\mathbf{R}}_{nn}(t,f)
=
\mathbf{L}(t,f)\mathbf{L}^{\mathrm{H}}(t,f).
\end{equation}

To ensure that the enhanced signal is invariant to this scaling, the mixture covariance used in the Wiener filter is normalized by the same factor,
\begin{equation}
\tilde{\mathbf{R}}_{xx}(t,f)
=
\frac{\mathbf{R}_{xx}(t,f)}{\gamma(f)},
\end{equation}
and both normalized covariance estimates are used in the DP-MWF in~\eqref{eq:DP_MWF}.

\subsection{Neural Network Architectures}
\subsubsection{OnlineSpatialNet}
We employ the OnlineSpatialNet architecture~\cite{Quan2024a} to estimate spatial covariance matrices from multichannel STFT observations. The network first projects the complex multichannel input at each time–frequency bin to a latent representation using a convolutional front-end, followed by interleaved narrow-band and cross-band processing blocks. The narrow-band blocks model temporal and spatial dynamics independently per frequency, while the cross-band blocks capture inter-frequency dependencies.
In the selected causal online variant, self-attention in the narrow-band blocks is replaced by a retention mechanism~\cite{Sun2023} that emphasizes recent frames while still incorporating long-term context, which is beneficial for tracking time-varying spatial statistics.

\subsubsection{NICE}
As a representative comparison, we use the NICE model~\cite{Casebeer2021a}, which applies a single-channel mask to the array signals and models temporal dynamics with a recurrent LSTM network.

To enable a fully blind comparison, we employ a shared single-channel mask estimator consisting of batch normalization, a unidirectional LSTM, and a linear projection, similar to~\cite{Heymann2016a}. The same network is applied independently to each microphone channel to produce per-channel masks, which are fused via a median operation to obtain a robust single-channel mask.

\section{Experimental Setup}
\subsection{Dataset}
We generate a dataset using speech and noise samples of the DNS challenge 4 dataset~\cite{Dubey2022} and combine them with a room acoustic and array simulation using pyroomacoustics~\cite{Scheibler2018}, similar to~\cite{Casebeer2021a}. 

Each scene is created by simulating room impulse responses (RIRs) for a 6-microphone circular array with \SI{7}{cm} diameter, with one speech source and one to three noise sources. Sources and array are randomly placed in shoebox rooms with dimensions uniformly sampled between \SI{4}{m} and \SI{8}{m} and reverberation times between \SI{0.25}{s} and \SI{0.75}{s}. The array position and height (\SIrange{1.2}{1.8}{m}) as well as source heights (\SIrange{1.0}{2.0}{m}) are randomized, enforcing a minimum distance of \SI{0.8}{m} between sources and the array.

Signals are convolved with the simulated array RIRs to generate \SI{5}{s} scenes, repeating shorter signals as needed. A random global SNR uniformly sampled between \SI{\pm 5}{\decibel} is applied based on the reverberant speech and summed noise signals. All data is resampled to \SI{32}{kHz}.

In total, the dataset comprises 30{,}000 scenes (\SI{41.7}{h}), split into training, validation, and test sets with an \SI{80}{\percent}/\SI{10}{\percent}/\SI{10}{\percent} ratio.

\subsection{Model Setup}
Both OnlineSpatialNet and NICE operate on \SI{32}{kHz} audio using an STFT with a frame size of 512 samples, hop size 256 samples, and a square-root Hann window. In both cases, the linear output layer is adapted to estimate the Cholesky factor (see Sec.~\ref{sec:cholesky-estim}). Enhancement is performed using the DP-MWF in~\eqref{eq:DP_MWF}, based on the noise covariance reconstructed from the estimated Cholesky factor and the mixture covariance obtained via causal averaging over a \SI{100}{ms} window. The DP-MWF parameters are set to $a=0$, $\mu=1$, and $\nu=8$.

The OnlineSpatialNet largely follows the configuration of~\cite{Quan2024a}, but in a \emph{tiny} configuration with reduced complexity, using 4 cross-band/narrow-band blocks, a hidden dimension of 64, and a T-ConvFFN with hidden dimension 128.

The mask estimator of NICE consists of a 3-layer unidirectional LSTM with hidden size 256 and dropout 0.2, followed by a linear layer with sigmoid activation. The covariance estimator employs a 2-layer unidirectional LSTM with hidden size 256 and frequency-sharing convolutional layers, following the best-performing but most complex configuration in~\cite{Casebeer2021a}.

As shown in Tab.~\ref{tab:model_complexity}, the selected configuration of OnlineSpatialNet requires only about one third of the parameters and computational cost, measured in giga floating-point operations per second (GFLOPs/s) of input audio, compared to NICE.
\begin{table}[t]
\centering
\caption{Model complexity comparison. For NICE, the parameter count is additionally broken down into the mask and covariance estimator.}
\label{tab:model_complexity}
\begin{tabular}{lcc}
\toprule
Model & Params [M] & GFLOPs/s \\
\midrule
OnlineSpatialNet & 0.82 & 23.23 \\
NICE & 2.54 (1.65+0.89) & 59.71 \\
\bottomrule
\end{tabular}
\end{table}

\subsection{Training Loss}
Both models are trained end-to-end with a combined loss consisting of a time-domain SI-SDR speech enhancement loss and a loss on the estimated noise Cholesky factor,
\begin{equation}
\mathcal{L}
=
\mathcal{L}_{\mathrm{SI\text{-}SDR}}
+
\lambda_{\mathrm{Chol}} \mathcal{L}_{\mathrm{Chol}},
\end{equation}
with $\lambda_{\mathrm{Chol}}=10$.
$\mathcal{L}_{\mathrm{SI\text{-}SDR}}$ is a multichannel SI-SDR loss where all microphone channels share the same scaling factor 
\begin{equation}
\alpha
=
\frac{\sum_t \mathbf{y}(t)^\top \mathbf{s}(t)}
{\sum_t \mathbf{s}(t)^\top \mathbf{s}(t)}
\end{equation}
to preserve inter-channel relations.
The SI-SDR loss is then defined as~\cite{Roux2019}
\begin{equation}
\mathcal{L}_{\mathrm{SI\text{-}SDR}}
=
-10 \log_{10}
\left(
\frac{\sum_t \|\alpha \mathbf{s}(t)\|_2^2}
{\sum_t \|\mathbf{y}(t) - \alpha \mathbf{s}(t)\|_2^2}
\right).
\end{equation}
To stabilize training, we use a normalized Frobenius loss between the predicted and reference Cholesky factors $\hat{\mathbf{L}}(t,f)$ and $\mathbf{L}(t,f)$,
\begin{equation}
\mathcal{L}_{\mathrm{Chol}}
=
\frac{\| \hat{\mathbf{L}}(t,f) - \mathbf{L}(t,f) \|_\mathrm{F}}
{\left\| \mathbf{L}(t,f) \right\|_\mathrm{F}},
\end{equation}
where the reference Cholesky factor was computed from the oracle noise-only covariance and $\|\cdot\|_\mathrm{F}$ denotes the Frobenius norm.

\begin{table*}[t]
\centering
\caption{Comparison of evaluation metrics.}
\label{tab:results}
\begin{tabular}{lcccccc}
\hline
Model & SI-SDR (dB) $\uparrow$ & $\mathcal{L}_{\mathrm{Chol}}$ $\downarrow$ & NR (dB) $\uparrow$ & CovSim $\uparrow$ & SpeechSim $\uparrow$ & NoiseSim $\uparrow$ \\
\hline
OnlineSpatialNet & \textbf{9.37} & \textbf{0.32} & 11.72 & \textbf{0.93} & \textbf{0.83} & 0.89 \\
NICE & 8.50 & 0.38 & \textbf{12.11} & 0.92 & 0.82 & \textbf{0.90} \\
\hline
Unprocessed & 0.02 & n/a & n/a & n/a & 0.79 & n/a \\
Oracle DP-MWF & 11.01 & n/a & 15.61 & n/a & 0.90 & 0.88 \\
\hline
\end{tabular}

\end{table*}

\subsection{Training}
Training uses Adam~\cite{Kingma2015} with learning rate $10^{-3}$ and gradient clipping with maximum norm 10. Both models are trained for 80 epochs, with the OnlineSpatialNet using a batch size of 4 and NICE a batch size of 8. The model checkpoint with the lowest validation loss is selected for evaluation.

\subsection{Evaluation Metrics}
Apart from the SI-SDR and Cholesky loss, we evaluate the direction-preserving multichannel enhancement systems using a set of metrics that quantify covariance estimation accuracy, spatial distortion, and noise reduction performance similar as in~\cite{Herzog2021a}.

To quantify noise reduction performance, we apply the estimated time-varying Wiener filter to the noise-only component and compare the input and output noise energies,
\begin{equation}
\mathrm{NR}
=
10 \log_{10}
\frac{\sum_{t,f} \|\mathbf{n}(t,f)\|_2^2}
{\sum_{t,f} \|\mathbf{W}(t,f)\mathbf{n}(t,f)\|_2^2}.
\end{equation}

To compare two complex covariance matrices $\mathbf{R}_1,\mathbf{R}_2\in\mathbb{C}^{M\times M}$, we use a cosine similarity measure
\begin{equation}
\sigma(\mathbf{R}_1,\mathbf{R}_2)
=
\frac{\Re\left\{\mathrm{tr}\!\left(\mathbf{R}_1^\mathrm{H}\mathbf{R}_2\right)\right\}}
{\|\mathbf{R}_1\|_\mathrm{F} \|\mathbf{R}_2\|_\mathrm{F}},
\end{equation}
which corresponds to the cosine similarity between vectorized covariance matrices. Here, $\Re\{\cdot\}$ denotes the real part, and $0 \leq \sigma(\mathbf{R}_1,\mathbf{R}_2) \leq 1$. 

To assess how well the estimated noise covariance matches the oracle noise covariance, we compute
\begin{equation}
\mathrm{CovSim}
=
\frac{1}{|\Omega|}
\sum_{(f,t)\in\Omega}
\sigma\!\left(
\widehat{\mathbf{R}}_{nn}(f,t),
\mathbf{R}_{nn}(f,t)
\right),
\end{equation}
where $\Omega$ is the set of all evaluated time-frequency covariance matrices.

To quantify how closely the covariance of the enhanced output $\mathbf{R}_{yy}(f,t)$ matches the spatial covariance structure of the clean speech $\mathbf{R}_{ss}(f,t)$, we further compute
\begin{equation}
\mathrm{SpeechSim}
=
\frac{1}{|\Omega|}
\sum_{(f,t)\in\Omega}
\sigma\!\left(
\mathbf{R}_{yy}(f,t),
\mathbf{R}_{ss}(f,t)
\right).
\end{equation}

Finally, we evaluate the spatial distortion when applying the estimated Wiener filter to the noise-only signal, yielding a filtered noise covariance $\bar{\mathbf{R}}_{nn}(f,t)$,
\begin{equation}
\mathrm{NoiseSim}
=
\frac{1}{|\Omega|}
\sum_{(f,t)\in\Omega}
\sigma\!\left(
\bar{\mathbf{R}}_{nn}(f,t),
\mathbf{R}_{nn}(f,t)
\right).
\end{equation}
These metrics measure how well spatial structure is preserved in the enhanced speech and residual noise, and thus quantify the spatial distortion introduced by the enhancement.

To evaluate performance in downstream applications, we steer a delay-and-sum beamformer toward the oracle target direction and measure the SI-SDR of the resulting signal. In addition, we perform binaural rendering using the end-to-end magnitude least squares method~\cite{Deppisch2021a} based on array transfer functions as in~\cite{Deppisch2024a}, and evaluate the interaural level difference (ILD) error. The ILD of a binaural signal is defined from the left- and right-ear energies as
\begin{equation}
\mathrm{ILD}
=
10 \log_{10}
\frac{\sum_t y_L^2(t)}
{\sum_t y_R^2(t)},
\end{equation}
and the corresponding ILD error is computed as the absolute difference between the estimated and reference ILD values, averaged over all test examples.


\section{Results}
Tab.~\ref{tab:results} summarizes the results for the OnlineSpatialNet and NICE configurations. The unprocessed mixture and an oracle DP-MWF based on the noise covariance calculated from the oracle noise-only signal via causal averaging are included as lower and upper performance bounds.

OnlineSpatialNet outperforms NICE in both SI-SDR and the Cholesky loss $\mathcal{L}_{\mathrm{Chol}}$, which were used as training objectives. With an SI-SDR of \SI{9.43}{\decibel}, it approaches the oracle DP-MWF (\SI{11.17}{\decibel}). NICE, however achieves slightly higher noise reduction, although it does not fully reach the performance of the oracle DP-MWF.

The improved covariance estimation accuracy is reflected in the lower $\mathcal{L}_{\mathrm{Chol}}$. In terms of covariance cosine similarity, both models achieve similar performance for all three metrics (CovSim, SpeechSim, and NoiseSim), suggesting that the directional alignment of noise covariances, and of enhanced speech and noise are similar. 

Overall, OnlineSpatialNet achieves strong speech enhancement and noise reduction while maintaining accurate spatial covariance estimates in terms of both normalized Frobenius error and cosine similarity. NICE provides slightly better noise reduction at the cost of higher speech distortion. Importantly, OnlineSpatialNet achieves these results with approximately one third of the parameters and computational cost (GFLOPs/s) of NICE (see Tab.~\ref{tab:model_complexity}).

\subsection{Downstream Applications}
\setlength{\tabcolsep}{4pt}
\begin{table}[t]
\caption{Evaluation metrics for downstream applications: beamformed delay-and-sum SI-SDR and binaural ILD error.}
\label{tab:downstream_metrics}
\centering
\begin{tabular}{lcc}
\hline
Model & DS SI-SDR (dB) $\uparrow$ & $\Delta$ILD (dB) $\downarrow$ \\
\hline
OnlineSpatialNet & \textbf{5.61} & \textbf{0.28} \\
NICE & 5.27 & 0.37 \\
\hline
Unprocessed & -0.11 & 1.27 \\
Oracle DP-MWF & 6.46 & 0.20 \\
\hline
\end{tabular}

\end{table}
Tab.~\ref{tab:downstream_metrics} reports the SI-SDR after delay-and-sum beamforming in the target direction, as well as the ILD error after binaural rendering. Consistent with the multichannel results, OnlineSpatialNet outperforms NICE in both metrics and approaches the performance of the oracle DP-MWF.

\begin{figure}[t]
    \centering
    \includegraphics[width=\linewidth]{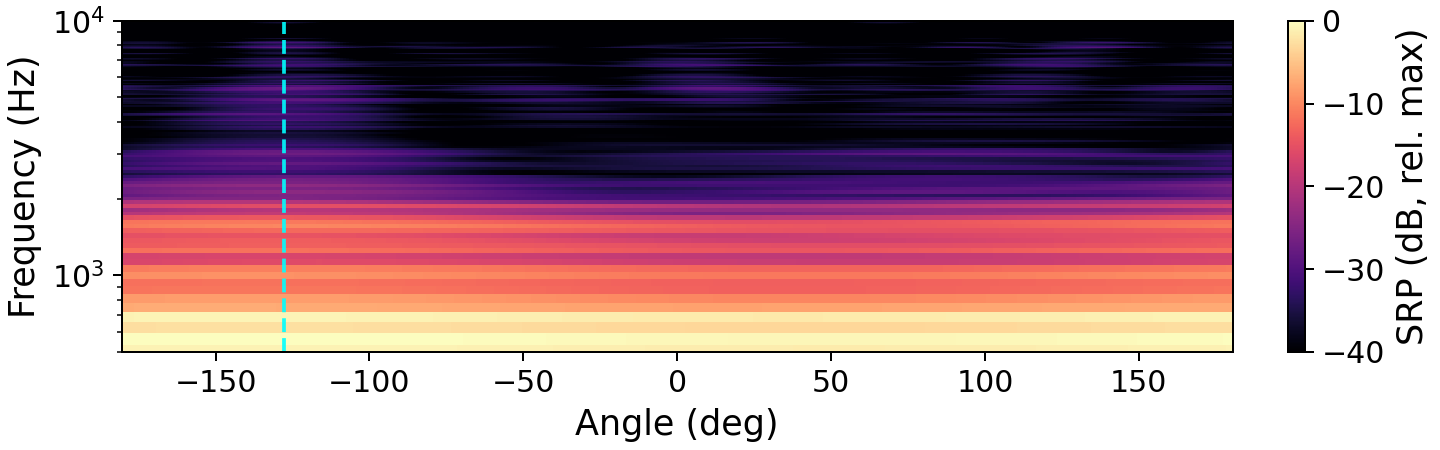}
    \put(-85,60){\footnotesize{\color{white}Clean Target}}\\
    \vspace{-0.5cm}
    \includegraphics[width=\linewidth]{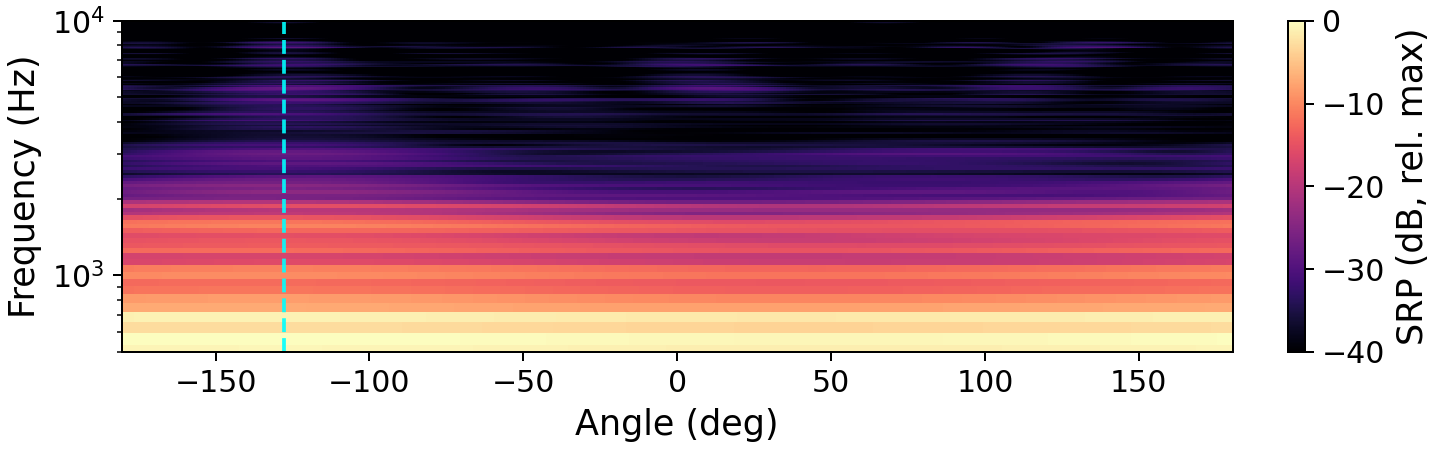}
    \put(-100,60){\footnotesize{\color{white}OnlineSpatialNet}}\\
    \vspace{-0.5cm}
    \includegraphics[width=\linewidth]{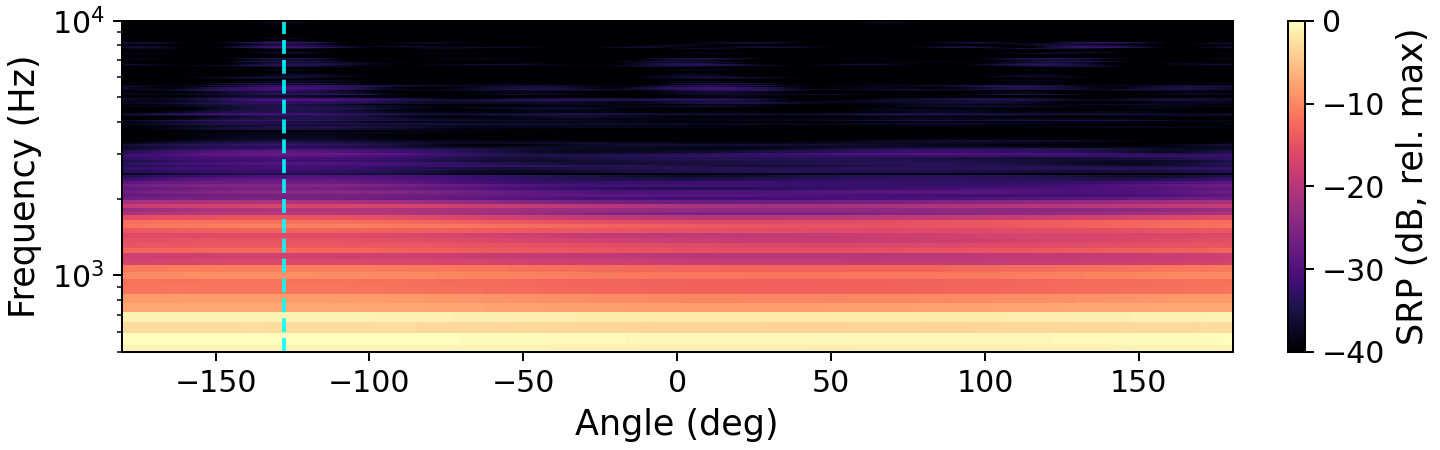}
    \put(-58,60){\footnotesize{\color{white}NICE}}\\
    \vspace{-0.5cm}
    \includegraphics[width=\linewidth]{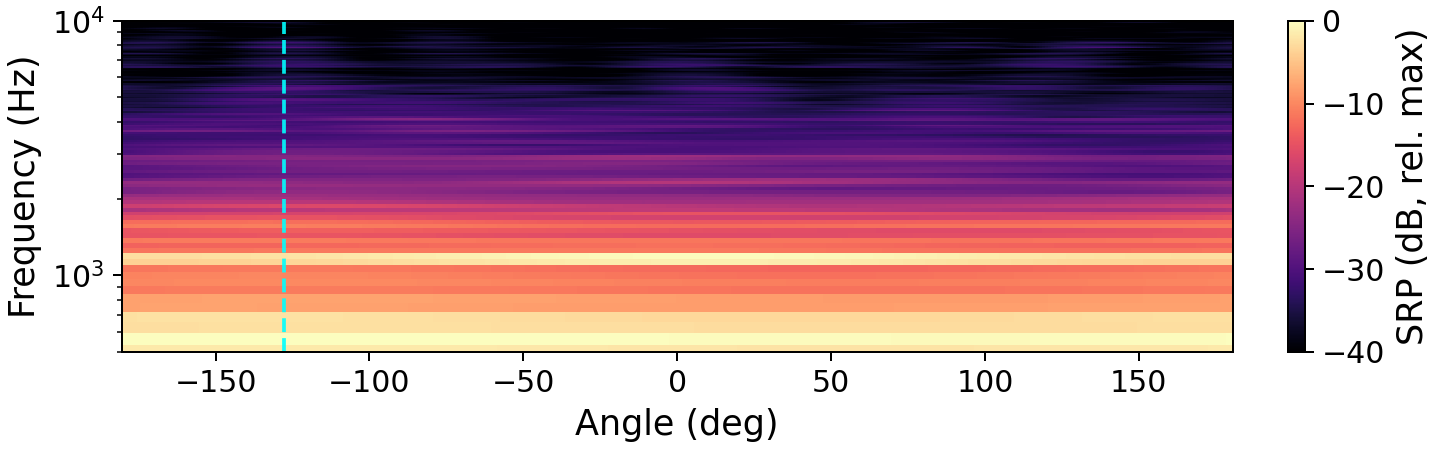}
    \put(-82,60){\footnotesize{\color{white}Unprocessed}}
    \caption{Steered response power maps of the clean target, the enhanced outputs of OnlineSpatialNet and NICE, and the unprocessed noisy mixture. The dashed vertical line indicates the target source direction.}
    \label{fig:SRP-maps}
\end{figure}

Fig.~\ref{fig:SRP-maps} shows steered response power maps over azimuth and frequency for an example scene. While the target direction is not clearly discernible in the noisy mixture, it becomes clearly visible after enhancement with both methods, with maps closely resembling that of the clean target signal.

Binaural audio examples are provided online\footnote{\url{https://thomasdeppisch.github.io/MIMO-speech-enhancement/}}.

\section{Conclusion}
We presented a direction-preserving MIMO speech enhancement approach based on neural covariance estimation. A lightweight OnlineSpatialNet is used to estimate the noise covariance, enabling end-to-end optimization of the enhancement process.

The results show that the proposed method, using the OnlineSpatialNet with only 820k parameters, yields improved speech enhancement and similar directional covariance alignment but slightly lower noise reduction performance compared to the more computationally demanding NICE model. Additional evaluations, including enhancement performance after beamforming and ILD error after binaural rendering, demonstrate its suitability for downstream spatial audio applications. Overall, the method provides a MIMO speech enhancement frontend for microphone arrays that preserves directional information while remaining computationally efficient. 

\section{Acknowledgment}
The computations were enabled by resources provided by the National Academic Infrastructure for Supercomputing in Sweden (NAISS), partially funded by the Swedish Research Council through grant agreement no. 2022-06725.

\bibliographystyle{IEEEtran}
\bibliography{references}

@inproceedings{Kingma2015,
    title = {{Adam: A method for stochastic optimization}},
    year = {2015},
    booktitle = {3rd Int. Conf. on Learning Representations (ICLR)},
    author = {Kingma, Diederik P. and Ba, Jimmy Lei},
    pages = {1--15}
}

@article{Herzog2023,
    title = {{AmbiSep: Joint Ambisonic-to-Ambisonic Speech Separation and Noise Reduction}},
    year = {2023},
    journal = {IEEE/ACM Transactions on Audio Speech and Language Processing},
    author = {Herzog, Adrian and Chetupalli, Srikanth Raj and Habets, Emanuël A.P.},
    pages = {3081--3094},
    volume = {31},
    doi = {10.1109/TASLP.2023.3297954}
}

@inproceedings{Wang2024c,
    title = {{Attention-Driven Multichannel Speech Enhancement in Moving Sound Source Scenarios}},
    year = {2024},
    booktitle = {IEEE Int. Conf. on Acoustics, Speech and Signal Processing (ICASSP)},
    author = {Wang, Yuzhu and Politis, Archontis and Virtanen, Tuomas},
    pages = {11221--11225},
    publisher = {IEEE},
    doi = {10.1109/ICASSP48485.2024.10448177}
}

@article{Deppisch2024a,
    title = {{Blind Identification of Binaural Room Impulse Responses from Smart Glasses}},
    year = {2024},
    journal = {IEEE/ACM Transactions on Audio, Speech, and Language Processing},
    author = {Deppisch, Thomas and Meyer-Kahlen, Nils and Amengual Gar{\'{i}}, Sebastià V.},
    pages = {4052–4065},
    volume = {32}
}

@inproceedings{Grinstein2025,
    title = {{Controlling the Parameterized Multi-channel Wiener Filter using a tiny neural network}},
    year = {2025},
    booktitle = {IEEE Workshop on Applications of Signal Processing to Audio and Acoustics},
    author = {Grinstein, Eric and Pandey, Ashutosh and Li, Cole and Srinivas, Shanmukha and Azcarreta, Juan and Donley, Jacob and Lee, Sanha and Aroudi, Ali and Bilen, Cagdas},
    doi = {10.1109/WASPAA66052.2025.11230930}
}

@inproceedings{Pandey2024a,
    title = {{Decoupled Spatial and Temporal Processing for Resource Efficient Multichannel Speech Enhancement}},
    year = {2024},
    booktitle = {IEEE Int. Conf. on Acoustics, Speech and Signal Processing (ICASSP)},
    author = {Pandey, Ashutosh and Xu, Buye},
    pages = {12206--12210}
}

@article{Herzog2020,
    title = {{Direction and reverberation preserving noise reduction of ambisonics signals}},
    year = {2020},
    journal = {IEEE/ACM Transactions on Audio Speech and Language Processing},
    author = {Herzog, Adrian and Habets, Emanuël A. P.},
    pages = {2461--2475},
    volume = {28},
    doi = {10.1109/TASLP.2020.3013979}
}

@inproceedings{Herzog2019e,
    title = {{Direction Preserving Wiener Matrix Filtering for Ambisonic Input-output Systems}},
    year = {2019},
    booktitle = {IEEE Int. Conf. on Acoustics, Speech and Signal Processing (ICASSP)},
    author = {Herzog, Adrian and Habets, Emanuel A.P.},
    pages = {446--450},
    doi = {10.1109/ICASSP.2019.8682873}
}

@inproceedings{Deppisch2021a,
    title = {{End-to-End Magnitude Least Squares Binaural Rendering of Spherical Microphone Array Signals}},
    year = {2021},
    booktitle = {Int. Conf. on Immersive and 3D Audio},
    author = {Deppisch, Thomas and Helmholz, Hannes and Ahrens, Jens},
    pages = {1--8}
}

@article{Doclo2002b,
    title = {{GSVD-based optimal filtering for single and multimicrophone speech enhancement}},
    year = {2002},
    journal = {IEEE Transactions on Signal Processing},
    author = {Doclo, Simon and Moonen, Marc},
    number = {9},
    pages = {2230--2244},
    volume = {50},
    doi = {10.1109/TSP.2002.801937}
}

@inproceedings{Dubey2022,
    title = {{ICASSP 2022 Deep Noise Suppression Challenge}},
    year = {2022},
    booktitle = {IEEE Int. Conf. on Acoustics, Speech and Signal Processing (ICASSP)},
    author = {Dubey, Harishchandra and Gopal, Vishak and Cutler, Ross and Aazami, Ashkan and Matusevych, Sergiy and Braun, Sebastian and Eskimez, Sefik Emre and Thakker, Manthan and Yoshioka, Takuya and Gamper, Hannes and Aichner, Robert},
    pages = {9271--9275}
}

@inproceedings{Erdogan2016,
    title = {{Improved MVDR beamforming using single-channel mask prediction networks}},
    year = {2016},
    booktitle = {Proc. INTERSPEECH},
    author = {Erdogan, Hakan and Hershey, John and Watanabe, Shinji and Mandel, Michael and Le Roux, Jonathan},
    pages = {1981--1985},
    doi = {10.21437/Interspeech.2016-552}
}

@book{Benesty2008,
    title = {{Microphone Array Signal Processing}},
    year = {2008},
    author = {Benesty, Jacob and Chen, Jingdong and Huang, Yiteng},
    publisher = {Springer Berlin, Heidelberg}
}

@article{Lugasi2024a,
    title = {{Multi-Channel to Multi-Channel Noise Reduction and Reverberant Speech Preservation in Time-Varying Acoustic Scenes for Binaural Reproduction}},
    year = {2024},
    journal = {IEEE/ACM Transactions on Audio Speech and Language Processing},
    author = {Lugasi, Moti and Donley, Jacob and Menon, Anjali and Tourbabin, Vladimir and Rafaely, Boaz},
    pages = {3283--3295},
    volume = {32},
    doi = {10.1109/TASLP.2024.3416668}
}

@article{Quan2024a,
    title = {{Multichannel Long-Term Streaming Neural Speech Enhancement for Static and Moving Speakers}},
    year = {2024},
    journal = {IEEE Signal Processing Letters},
    author = {Quan, Changsheng and Li, Xiaofei},
    number = {8},
    pages = {2295--2299},
    volume = {31},
    doi = {10.1109/LSP.2024.3418714}
}

@inproceedings{Heymann2016a,
    title = {{Neural Network Based Spectral Mask Estimation for Acoustic Beamforming}},
    year = {2016},
    booktitle = {IEEE Int. Conf. on Acoustics, Speech, and Signal Processing (ICASSP)},
    author = {Heymann, Jahn and Drude, Lukas and Haeb-Umback, Reinhold},
    pages = {196--200},
    isbn = {9781479999880}
}

@article{Casebeer2021a,
    title = {{NICE-Beam: Neural Integrated Covariance Estimators for Time-Varying Beamformers}},
    year = {2021},
    journal = {arXiv:2112.04613},
    author = {Casebeer, Jonah and Donley, Jacob and Wong, Daniel and Xu, Buye and Kumar, Anurag},
    arxivId = {2112.04613}
}

@article{Souden2010,
    title = {{On optimal frequency-domain multichannel linear filtering for noise reduction}},
    year = {2010},
    journal = {IEEE Transactions on Audio, Speech and Language Processing},
    author = {Souden, Mehrez and Benesty, Jacob and Affes, Sofiène},
    number = {2},
    pages = {260--276},
    volume = {18},
    doi = {10.1109/TASL.2009.2025790}
}

@inproceedings{Scheibler2018,
    title = {{Pyroomacoustics: A Python package for audio room simulations and array processing algorithms}},
    year = {2018},
    booktitle = {IEEE Int. Conf. on Acoustics, Speech and Signal Processing (ICASSP)},
    author = {Scheibler, Robin and Bezzam, Eric and Dokmanic, Ivan},
    pages = {351–355}
}

@article{Sun2023,
    title = {{Retentive Network: A Successor to Transformer for Large Language Models}},
    year = {2023},
    journal = {arXiv:2307.08621},
    author = {Sun, Yutao and Dong, Li and Huang, Shaohan and Ma, Shuming and Xia, Yuqing and Xue, Jilong and Wang, Jianyong and Wei, Furu},
    pages = {1--14},
    arxivId = {2307.08621}
}

@inproceedings{Roux2019,
    title = {{SDR - Half-baked or Well Done?}},
    year = {2019},
    booktitle = {IEEE Int. Conf. on Acoustics, Speech and Signal Processing (ICASSP)},
    author = {Roux, Jonathan Le and Wisdom, Scott and Erdogan, Hakan and Hershey, John R.},
    pages = {626--630}
}

@inproceedings{Herzog2021a,
    title = {{Signal-Dependent Mixing for Direction-Preserving Multichannel Noise Reduction}},
    year = {2021},
    booktitle = {29th European Signal Processing Conference},
    author = {Herzog, Adrian and Habets, Emanuël A. P.},
    pages = {96--100},
    doi = {10.23919/EUSIPCO54536.2021.9616236}
}

@article{Spriet2004,
    title = {{Spatially pre-processed speech distortion weighted multi-channel Wiener filtering for noise reduction}},
    year = {2004},
    journal = {Signal Processing},
    author = {Spriet, A. and Moonen, M. and Wouters, J.},
    number = {12},
    pages = {2367--2387},
    volume = {84},
    doi = {10.1016/j.sigpro.2004.07.028}
}

@article{Lugasi2020,
    title = {{Speech Enhancement Using Masking for Binaural Reproduction of Ambisonics Signals}},
    year = {2020},
    journal = {IEEE/ACM Transactions on Audio Speech and Language Processing},
    author = {Lugasi, Moti and Rafaely, Boaz},
    pages = {1767--1777},
    volume = {28},
    doi = {10.1109/TASLP.2020.2998294}
}

\end{document}